\documentclass[twocolumn,trackchanges]{aastex63}

\usepackage{mathtools,amssymb}
\usepackage{booktabs}
\shorttitle{Multi-Wavelength Characterization of Rings in HD 169142}
\shortauthors{Mac\'{\i}as et al.}

\begin{document}


\title{Characterization of Ring Substructures in the Protoplanetary Disk of HD 169142 from Multi-Wavelength ALMA Observations}


\author{Enrique Mac\'{\i}as} 
\affil{Department of Astronomy, Boston University, 725 Commonwealth Avenue, Boston, MA 02215, USA
{\tt emacias@bu.edu}}

\author{Catherine C. Espaillat}
\affil{Department of Astronomy, Boston University, 725 Commonwealth Avenue, Boston, MA 02215, USA
{\tt emacias@bu.edu}}
\author{Mayra Osorio}
\affil{Instituto de Astrof\'\i sica de Andaluc\'\i a (CSIC) Glorieta de la Astronom\'\i a s/n E-18008 Granada, Spain}
\author{Guillem Anglada}
\affil{Instituto de Astrof\'\i sica de Andaluc\'\i a (CSIC) Glorieta de la Astronom\'\i a s/n E-18008 Granada, Spain}
\author{Jos\'e M. Torrelles}
\affil{Institut de Ci\`encies de l'Espai (CSIC) and Institut d'Estudis Espacials de Catalunya (IEEC), Can Magrans S/N, Cerdanyola del Vall\`es (Barcelona), Spain}
\author{Carlos Carrasco-Gonz\'alez}
\affil{Instituto de Radioastronom\'{\i}a y Astrof\'{\i}sica UNAM, Apartado Postal 3-72 (Xangari), 58089 Morelia, Michoac\'an, Mexico}
\author{Mario Flock}
\affil{Max-Planck Institute for Astronomy, K\"onigstuhl 17, 69117, Heidelberg, Germany}
\author{Hendrik Linz} 
\affil{Max-Planck Institute for Astronomy, K\"onigstuhl 17, 69117, Heidelberg, Germany}
\author{Gesa H.-M. Bertrang}
\affil{Max-Planck Institute for Astronomy, K\"onigstuhl 17, 69117, Heidelberg, Germany}
\author{Thomas Henning}
\affil{Max-Planck Institute for Astronomy, K\"onigstuhl 17, 69117, Heidelberg, Germany}
\author{Jos\'e F. G\'omez}
\affil{Instituto de Astrof\'\i sica de Andaluc\'\i a (CSIC) Glorieta de la Astronom\'\i a s/n E-18008 Granada, Spain}
\author{Nuria Calvet}
\affil{Department of Astronomy, University of Michigan, 1085 South University Avenue, Ann Arbor, MI4810, USA}
\author{William R. F. Dent}
\affil{ALMA JAO, Alonso de Cordova 3107, Santiago, Chile}

\begin{abstract}
We present a detailed multi-wavelength characterization of the multi-ring disk of HD 169142. We report new ALMA observations at 3 mm and analyze them together with archival 0.89 and 1.3 mm data. Our observations resolve three out of the four rings in the disk previously seen in high-resolution ALMA data. A simple parametric model is used to estimate the radial profile of the dust optical depth, temperature, density, and particle size distribution. We find that the multiple ring features of the disk are produced by annular accumulations of large particles, probably associated with gas pressure bumps. Our model indicates that the maximum dust grain size in the rings is $\sim1$ cm, with slightly flatter power-law size distributions than the ISM-like size distribution ($p\sim3.5$) found in the gaps. In particular, the inner ring ($\sim26$ au) is associated with a strong and narrow buildup of dust particles that could harbor the necessary conditions to trigger the streaming instability. According to our analysis, the snowlines of the most important volatiles do not coincide with the observed substructures. We explore different ring formation mechanisms and find that planet-disk interactions are the most likely scenario to explain the main features of HD 169142. Overall, our multi-wavelength analysis provides some of the first unambiguous evidence of the presence of radial dust traps in the rings of HD 169142. A similar analysis in a larger sample of disks could provide key insights on the impact that disk substructures have on the dust evolution and planet formation processes.

\end{abstract}

\keywords{protoplanetary disks --- planet-disk interactions --- stars: individual (HD 169142) --- stars: pre-main sequence --- techniques: interferometric}



\section{Introduction} \label{sec:intro}

One of the most important recent discoveries in the planet formation field is the ubiquity of protoplanetary disk substructures \citep{lon18,and18a}. The implications of this discovery are  transformational, but still not fully understood. Since the first discoveries of the dominant presence of disk substructures \citep[e.g.,][]{oso14,alm15,and16,per16,ave18}, several possible origins have been proposed. The most likely and promising scenario is the dynamic interaction between the disk and one or more young planets \citep[e.g.,][]{zhu14,bae17,zha18}, yet other processes could still play an important role \citep[e.g.,][]{flo15,oku16,pin17}.

An important piece of information to understand the origin and role of disk substructures is their dust content. Various substructure-forming physical processes involve the onset of gas pressure bumps, which can trap and accumulate large dust particles into annular \citep[e.g.,][]{pin12} or azimuthally asymmetric structures \citep[e.g.,][]{bir13,lyr13}. Such non-smooth gas distributions have been proposed to be a key piece of the dust evolution and planet formation process, since they can stop the radial drift of large particles and allow them to grow up to planetesimal sizes \citep{whi72,bar95,bra08}.
Therefore, analyzing the dust particle size distribution in disks and in their substructures is crucial not only to discern between different substructure origins, but also to understand the role that these features play in the planet formation process.

One of the best methods to study the dust size distribution in disks is by analyzing the spectral behavior of the (sub-)mm dust continuum emission \citep[e.g.,][]{per15,taz16}. The spectral index ($\alpha$) at these wavelengths depends on the optical depth of the emission ($\tau_{\nu}$) and on  the power-law index ($\beta$) of the dust opacity (i.e., $\kappa_{\nu} \propto \nu^{\beta}$; \citealp{bec90}). If the dust emission is optically thin and in the Rayleigh Jeans regime, the spectral index will simply be $\alpha = 2+\beta$. The parameter $\beta$ depends in turn on the size distribution and maximum grain size of the dust: $\beta \sim 1.6-1.8$ is expected for micron-sized grains, while $\beta\sim0-1$ for dust populations dominated by sizes larger than mm/cm \citep{dal01}. Therefore, by analyzing the spectral index of optically thin dust emission, one can in principle study the size distribution of dust particles throughout the disk. 

Some recent studies using ALMA observations between 0.88 mm and 2 mm have started to find evidence of significant radial changes in the spectral indices of disks, with higher values in annular gaps and lower values in the ring substructures \citep{tsu16,hua18a}. These trends could hint at spatial changes in the dust size distribution, but they could also be the result of higher optical depths at the position of the rings. In order to discern between these two effects, observations at multiple wavelengths are necessary, including longer wavelengths where the dust emission will be optically thinner \citep[e.g.,][]{car16,liu17,mac18}.

In this paper we study the protoplanetary disk around the nearby (d=$113.6\pm0.8$ pc; \citealp{gai18,bai18}) Herbig Ae star HD 169142 (A8 Ve, $M_{\star}\simeq1.65~M_{\odot}$, age$\simeq10$ Myr; \citealp{car18}). This star is surrounded by an almost face-on ($i\simeq13^{\circ}$; \citealp{ram06,pan08}) pre-transitional disk \citep{mee10,hon12,esp14,oso14}, that has been extensively studied at multiple wavelengths.
The disk shows a $\sim$20 au inner cavity and two bright rings of emission, with radii $\sim25$ and $\sim60$ au. This multi-ring morphology was first inferred through near-IR polarimetric observations \citep{qua13}, and later confirmed to be associated with annular rings in the dust surface density of the disk \citep{oso14}. Later studies have analyzed the disk at near-IR \citep{lig18,mon17,poh17,ber18} and mm wavelengths \citep{fed17,mac17}, showing that the inner cavity and gap are not only depleted from mm-sized dust particles, but also from micron-sized grains and gas. Based on these observations, it has been proposed that the disk harbors two or more giant planets that are creating the signature double-ring morphology of HD 169142 \citep{fed17,ber18,car18}. \citet{reg14} reported the detection of a planet candidate inside the inner cavity through IR imaging, but the confirmation of this source has remained elusive, and it has been suggested that this feature might be instead associated with the inner ring \citep{bil14,lig18}. Multi-epoch VLT/SPHERE observations of HD 169142 have  suggested the presence of faint spiral arms in the disk, as well as a planet candidate inside the disk gap \citep{gra19}. More recently, \citep{per19} reported high angular resolution ALMA observations that resolved the outer ring into three separate rings. These authors proposed that a mini-Neptune located in the middle of this triple-ring could be the responsible for the formation of this feature. Overall, HD 169142 shows strong evidence of the presence of multiple young planets, but the origin of the disk substructures has not been fully confirmed yet.

Here we present a multi-wavelength analysis of the multi-ring disk of HD 169142 using ALMA observations at 0.89 mm, 1.3 mm, and 3 mm. In section 2 we report the observations, their calibration, and the imaging details. Section 3 outlines the results from the ALMA cleaned images. In section 4 we present our modeling approach of the observed visibilities, and estimate the underlying dust size distribution of the disk. Section 5 includes a discussion on the inferred optical depths and dust size distribution, and how this can help us understand the origin of the ring substructures. We summarize and conclude our analysis in section 6.

\section{Observations} \label{sec:observations}

We present new ALMA observations obtained at Band 3 (project code: 2016.1.01158.S), as well as archival ALMA data at Band 7 (project code: 2012.1.00799.S), and Band 6 (project codes: 2015.1.00490.S, and 2015.1.01301.S). The Band 7 observations were first reported in \citet{ber18}, but here we present a more detailed reduction of the data that allowed us to obtain a higher sensitivity and angular resolution. Details of the observations and setups are summarized in Table 1. The raw data were calibrated using the reduction scripts provided in the ALMA archive and their corresponding version of \texttt{CASA} (Common Astronomy Software Applications). Further calibration and imaging was performed in \texttt{CASA} version 5.3.0.

\floattable
\begin{deluxetable}{cccccccccc}
\tabletypesize{\scriptsize}
\tablecaption{Summary of ALMA Observations \label{tab:Obs}}
\tablehead{
\colhead{Project} &\colhead{P.I.}&\colhead{Date}&\colhead{On-source} &\colhead{$N_{\rm ant}$}&\colhead{Baselines} &\colhead{Freq. range}&\colhead{Flux} &\colhead{Bandpass} &\colhead{Phase}\\
\colhead{Code}& \colhead{}&\colhead{}& \colhead{time (min)}&\colhead{} & \colhead{(m)} &\colhead{(GHz)} & \colhead{cal.} & \colhead{cal.} & \colhead{cal.}}
\startdata
\multicolumn{10}{c}{Band 7} \\
\hline
\dataset[2012.1.00799.S]{https://almascience.nrao.edu/aq/?project\_code=2012.1.00799.S} & M. Honda & 2015 July 26 & 41.4 & 41 & 15\textendash1574 & 331.040\textendash332.915 &  J1924-2914 & J1924-2914 & J1826-2924 \\
 & & & & & & 343.050\textendash344.925 & & & \\
 & & 2015 July 27 & 21.2 & 41 & 15\textendash1574 & 331.040\textendash332.915 & J1924-2914 & J1924-2914 & J1826-2924 \\
  & & & & & & 343.050\textendash344.925 & & & \\
 & & 2015 08 Aug & 41.4 & 43 & 35\textendash1574 & 331.040\textendash332.915 &  Pallas & J1924-2914 & J1826-2924 \\
  & & & & & & 343.050\textendash344.925 & & & \\
\hline
\multicolumn{10}{c}{Band 6} \\
\hline
\dataset[2015.1.00490.S]{https://almascience.nrao.edu/aq/?project\_code=2015.1.00490.S} & M. Honda & 2016 September 14 & 49.6 & 38 & 15\textendash3200 & 232.029\textendash233.904& J1733-1304 & J1924-2914 & J1820-2528 \\
& & 2016 September 14 & 49.6 & 38 & 15\textendash3200 & 232.029\textendash233.904 & J1924-2914 & J1924-2914 & J1820-2528 \\
\dataset[2015.1.01301.S]{https://almascience.nrao.edu/aq/?project\_code=2015.1.01301.S} & J. Hashimoto & 2016 September 17 & 29.3 & 40 & 15\textendash3100 & 231.472\textendash233.472 & J1924-2914 & J1924-2914 & J1820-2528 \\
\hline
\multicolumn{10}{c}{Band 3} \\
\hline
\dataset[2016.1.01158.S]{https://almascience.nrao.edu/aq/?project\_code=2016.1.01158.S} & M. Osorio & 2017 May 07 & 7.8 & 40 & 15\textendash1124 & 89.495\textendash93.495 & J1924-2914 & J1924-2914 & J1826-2924  \\
 & & & & & & 101.495\textendash105.495 & \\
& & 2017 September 09 & 25.4 & 40 & 41\textendash7552 & 89.495\textendash93.495 & J1924-2914 & J1924-2914 & J1826-2924 \\
 & & & & & & 101.495\textendash105.495 & \\
\enddata
\end{deluxetable}

After inspecting the calibrated visibilities, each dataset was separately self-calibrated with the continuum emission. Phase self-calibration was first performed iteratively decreasing the solution interval in each step, from 120 s to 10 s, or until the peak signal-to-noise ratio (SNR) did not improve from the previous iteration. The \texttt{hogbom} algorithm within the \texttt{tclean} task was used for the imaging during the phase-only self-calibration, together with natural weighting. Afterwards, amplitude self-calibration was performed using the multi-term, multi-frequency synthesis algorithm (\texttt{mtmfs}; \citealp{rau11}) in \texttt{tclean}, assuming a linear spectrum ($nterms=2$) and point-like components ($scales=0$), as well as using natural weighting. Various iterations of amplitude self-calibration were attempted, performing at least one first iteration with a solution interval as long as the scan. In general, self-calibration resulted in substantial improvements of the peak SNR, typically by factors between 5 and 10. The only exception was the Band 3 data, which due to the much lower brightness temperature of the dust emission, resulted in improvements of $20\%$ and $10\%$ for the compact and extended datasets, respectively.

\begin{figure*}
\centering
\figurenum{1}
\includegraphics[width=\textwidth]{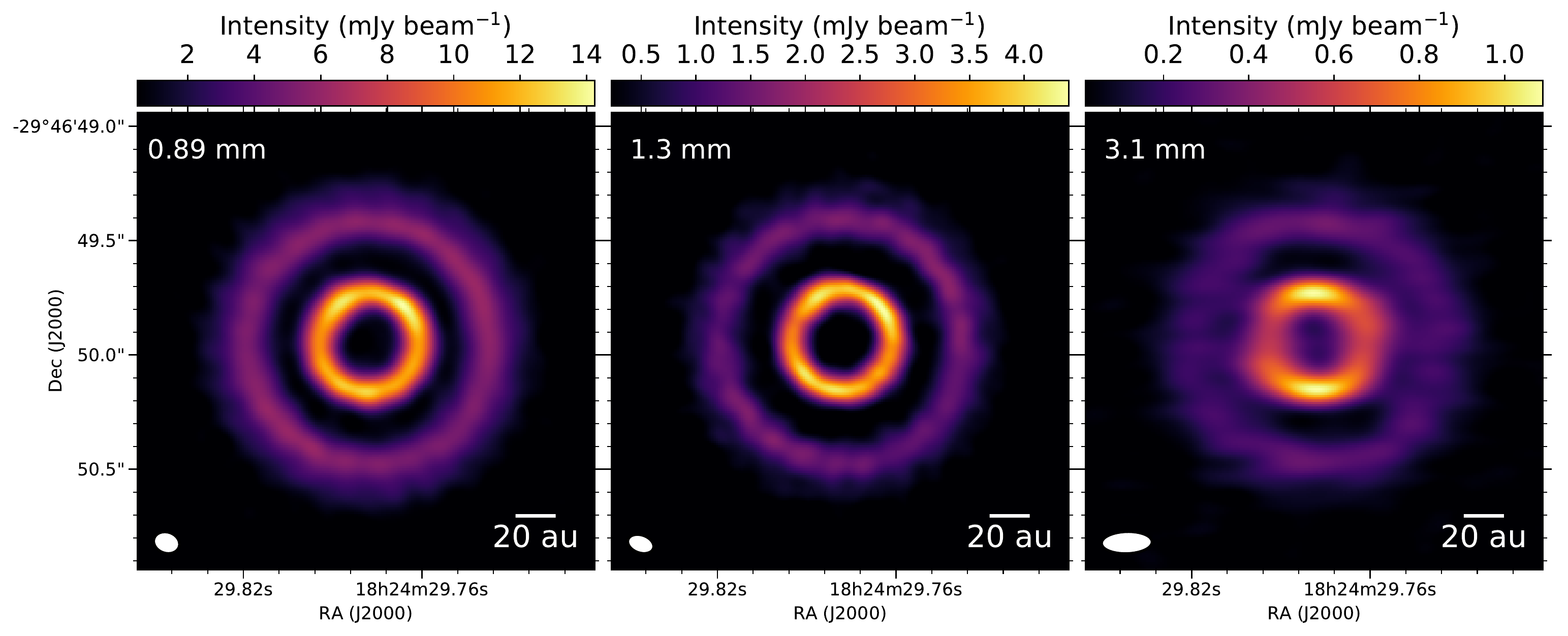}
\label{fig:maps}
\caption{Continuum emission of the multi-ring disk of HD 169142 at 0.89 mm, 1.3 mm, and 3 mm (from left to right). The beam size and rms sensitivity of each panel are listed in Table \ref{Tab:ObsRes}.}
\end{figure*}

In order to combine and compare the observations, all datasets were corrected for their proper motions and shifted to the first epoch of the Band 6 observations (2016 Sep 14). To do this, the \texttt{CASA} tasks \texttt{fixvis} and \texttt{fixplanets} were used, in combination with the position and proper motions measured by Gaia \citep{gai18}. Images were then obtained at 0.89 mm, 1.3 mm, and 3 mm by combining all the datasets at each band, and using the \texttt{mtmfs} algorithm with $nterms=2$ and multiple scales at 0, 1, 3, and 5 times the beam size. Different visibility weightings were tested to reach a compromise between image quality and angular resolution. The final images presented here (Fig. \ref{fig:maps}) were obtained with uniform weighting for Band 7 and 6, and Briggs weighting (robust$=0.0$) for Band 3. The resulting rms sensitivity, angular resolution, and integrated flux density at each band are listed in Table \ref{Tab:ObsRes}.

\floattable
\begin{deluxetable}{cccccc}
\tablecaption{Observation Results \label{Tab:ObsRes}}
\tablehead{
\colhead{Band} & \colhead{Central frequency} & \colhead{Wavelength} & \colhead{Beam shape}  & \colhead{rms} & \colhead{Flux Density\tablenotemark{a}}  \\
\colhead{} & \colhead{(GHz)} & \colhead{(mm)} & \colhead{} &  \colhead{(mJy beam$^{-1}$)} & \colhead{(mJy)}}
\startdata
7 & 338.184 &  0.886 & $0\rlap.''12\times0\rlap.''09$, PA$=71^{\circ}$ & 0.16 & $554\pm2$ \\
6+7 & 288.755 & 1.04 & $0\rlap.''11\times0\rlap.''08$, PA$=63^{\circ}$& 0.13 & $ 369\pm2$ \\
6 & 225.490 & 1.33 & $0\rlap.''12\times0\rlap.''08$, PA$=71^{\circ}$& 0.08 & $ 166.2\pm1.3$ \\
3 & 97.493   & 3.08 & $0\rlap.''22\times0\rlap.''10$, PA$=-88^{\circ}$ & 0.017 & $18.29\pm0.17$ \\
\enddata
\tablenotetext{a}{Flux density integrated within a circle $0\rlap.''9$ in radius centered on HD 169142.}
\end{deluxetable}

\begin{figure*}[t]
\centering
\figurenum{2}
\includegraphics[width=\textwidth]{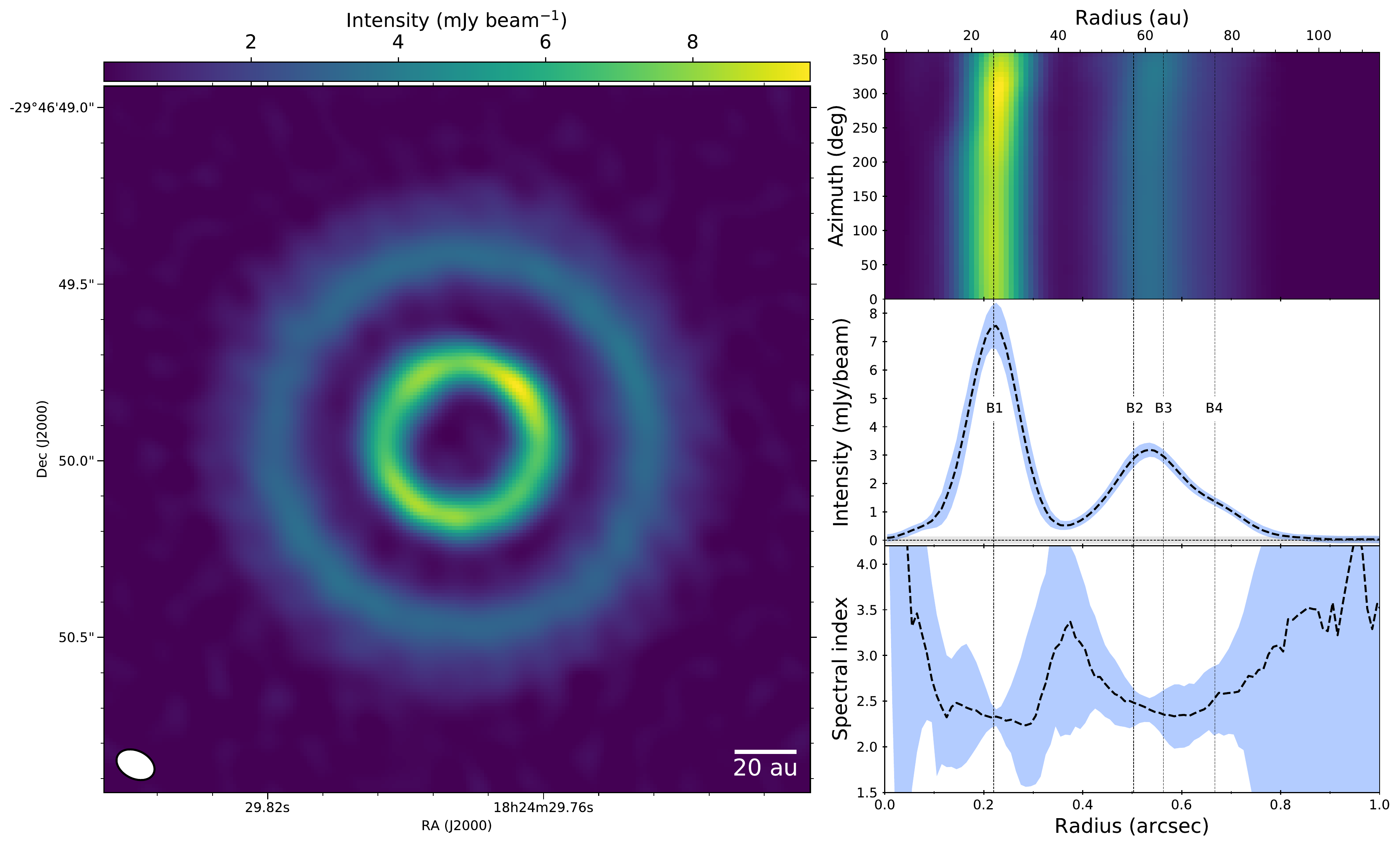}
\label{fig:MFS_map}
\caption{\textit{Left:} Image of the continuum emission of HD 169142 at $\sim1.04$ mm, obtained by combining the Band 6 and 7 data. Its beam size and rms sensitivity are listed in Table \ref{Tab:ObsRes}. \textit{Top-right:} Deprojection of the 1.04 mm image from the left panel, in polar coordinates. The azimuth angle is defined from N to E. The center of the deprojection is located at the position of the star determined by Gaia at the first epoch of the Band 6 observations (see \S\ref{sec:observations}). \textit{Middle-right:} Azimuthally averaged intensity profile of the 1.04 mm emission of HD 169142. The blue shaded region indicates the standard deviation of the emission at each radius. The horizontal line represents the y=0 level. The gray region around this line shows the $\pm1\sigma$ (rms of the image) level. \textit{Bottom-right:} Deprojected radial profile of the spectral index map between 0.89 mm and 1.3 mm. The dashed line shows the median of the spectral index at each radius, while the blue shaded region indicates the 16th and 84th percentile. The dashed vertical lines in the three right panels indicate the positions of the 4 rings detected by \citet{per19}.}
\end{figure*}

Finally, the Band 7 and Band 6 data were combined and imaged together in \texttt{tclean} using \texttt{mtmfs} with the same setup as for the individual images. By combining these two datasets, the cleaning algorithm is capable of producing a map of the spectral index $\alpha$. At the same time, the wider frequency extent of the combined datasets provides a larger uv-coverage, which results in a higher quality image at an intermediate frequency of 289 GHz. The resulting image is shown in Fig. \ref{fig:MFS_map}.  We attempted to obtain a similar image combining the Band 6 and Band 3 data, but the resulting image quality was substantially poorer, probably due to the larger separation in frequency and the significant difference in uv coverages.

\section{Observational Results} \label{sec:results}

The multi-ring morphology of the dust emission of HD 169142 is clearly resolved from 0.89 mm to 3.1 mm. Two ring-like components are detected peaking at radii $\sim0\rlap.''22$ ($\sim25$ au) and $\sim0\rlap.''53$ ($\sim60$ au), consistent with previous observations \citep[e.g.][]{oso14,fed17,mac17,car18,ber18,per19}. The 0.89 mm and 1.3 mm images, with a similar spatial resolution, successfully resolve the outer component in the radial direction, but only do it marginally for the inner one. Our 3.1 mm observations, on the other hand, display a very elongated beam that significantly distorts the emission of the inner component. 

At the spatial resolution of our observations (Table \ref{Tab:ObsRes}), the disk appears fairly axisymmetric. We note that some small azimuthal asymmetries can be seen in the inner ring at 0.89 mm and 1.3 mm, which are consistent with previous SPHERE \citep{ber18,gra19}, 7 mm VLA \citep{mac17}, and very high resolution 1.3 mm ALMA observations \citep{per19}. These asymmetries are consistent with tidal interactions with one or more massive planets in the inner and outer gaps of the disk \citep{ber18}. Nevertheless, they are just marginally resolved in our images, which makes it difficult to analyze them. Therefore, in the following we will focus on the axisymmetric features of HD 169142, and leave the multi-wavelength analysis of these asymmetries for future higher spatial resolution observations.

\begin{figure*}
\centering
\figurenum{3}
\includegraphics[width=\textwidth]{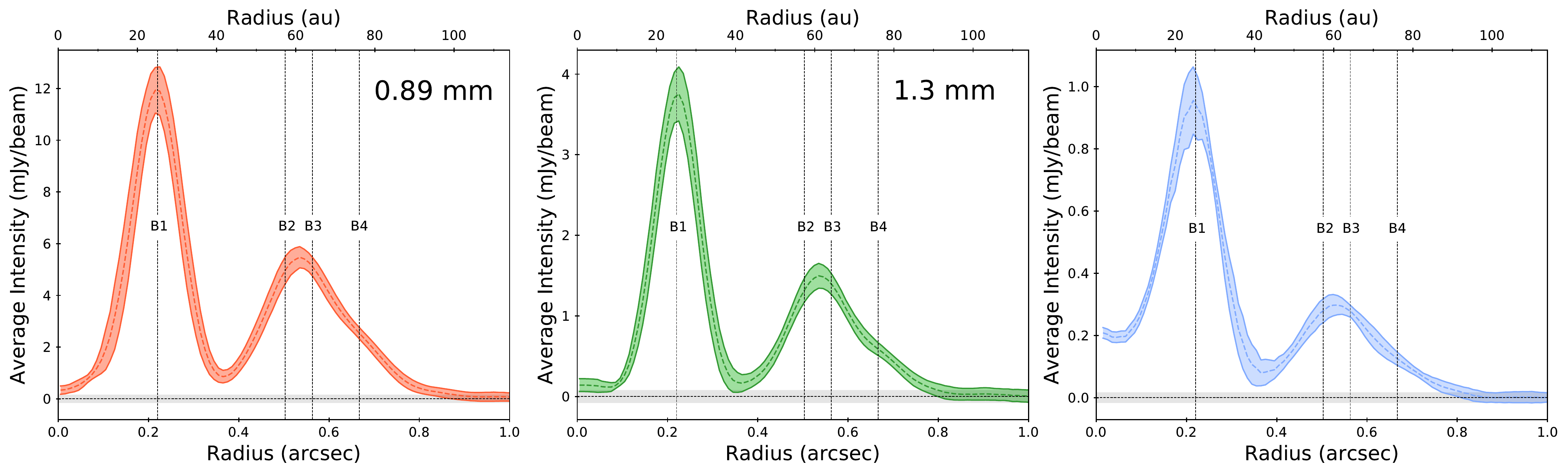}
\label{fig:profiles}
\caption{Radial profiles of continuum emission of the multi-ring disk of HD 169142 at 0.89 mm, 1.3 mm, and 3 mm (from left to right). The colored region indicates the standard deviation of the emission at each radius. 
The vertical lines show the positions of the 4 rings detected by \citet{per19}. The horizontal lines and gray regions around them represent the y=0 level and the $\pm1\sigma$ (rms of the images) level. }
\end{figure*}

The averaged radial intensity profiles of the three images are shown in Fig. \ref{fig:profiles}. These profiles were obtained by averaging the emission within concentric ellipses, projected with the inclination and PA of the disk ($i=13^{\circ}$, PA$=5^{\circ}$; \citealp{ram06}). In order to minimize the effects of the elongated beam at 3.1 mm, the profile at this band is calculated using only two $90^{\circ}$ cones to the North and South, along the minor axis of the beam. The profiles clearly show that the outer disk is resolved into a bright ring at $\sim0\rlap.''53$ ($\sim60$ au) followed by some extended emission that could be associated with an additional ring. The 0.89 mm and 1.3 mm emission display a small decrease of the emission in between these two components at $\sim0\rlap.''65$ ($\sim74$ au). Recent high spatial resolution observations at 1.3 mm have in fact resolved the outer component of the disk into three separate rings at $0\rlap.''503$, $0\rlap.''563$, and $0\rlap.''667$ \citep{per19}. The small dip in our profiles is associated with the gap between the last two rings in those observations, but the second and third rings remain unresolved in our images and appear as one single ring at an intermediate radius. VLT/SPHERE polarimetric observations at 0.73 $\mu$m and 1.2 $\mu$m also detected an additional annular ring at $\sim90$ au \citep{poh17,ber18}, which may also be seen at 7 mm \citep{mac17}. However, our ALMA observations do not detect any emission at this radius. 

In order to avoid confusion, in the following we will follow the notation by \citet{per19}  and we will refer to the disk substructures as ``B" (for bright annular substructures, a.k.a. rings) or ``D" (for dark annular substructures, a.k.a. gaps) followed by their rank in radial distance. In this way, we refer to the inner ring as B1, and to the substructures in the outer disk as B2, B3, and B4, which are all separated between each other by D1 (the inner cavity within B1), D2 (the gap between B1 and B2), D3 (between B2 and B3), and D4 (between B3 and B4). Since we do not resolve B2 from B3, we will refer to the combined bright ring at $\sim60$ au as B2/3. 

The image resulting from the combination of Band 6 and 7 data shows a similar morphology to the images in each band separately (see Fig. \ref{fig:MFS_map}). However, this combination now allows us to analyze the variations of the 0.89-1.3 mm spectral index ($\alpha$) throughout the disk. The bottom right panel of Fig. \ref{fig:MFS_map} shows the radial profile of the spectral index map between these two wavelengths. Due to the relatively high spread of values in some zones of the disk, we compute this profile with the median within each projected ellipse. The colored region in this case indicates the 16th and 84th percentiles. We note that we do not include in this figure the systematic error introduced by the absolute flux calibration at each band ($\sim10\%$ and $\sim5\%$ at Band 7 and 6, respectively). This uncertainty would shift the profile upwards or downwards, but its relative shape would be unaffected. As can be seen, $\alpha$ varies significantly throughout the disk. The $\alpha$ profile is relatively flat around $\sim2.5$ at the position of the inner ring and the outer disk. Even though the spread of $\alpha$ is higher in the gaps, the profile indicates that its value is higher in these regions.

This behavior has been found in other protoplanetary disks \citep[e.g.][]{tsu16,hua18a}, and it might indicate substantial changes in the size distribution of dust particles in the disk. The opacity law of small dust particles is usually steeper (i.e., with higher $\beta$, where $\kappa_{\nu}\propto \nu^{\beta}$) than that of large grains, so when its emission is optically thin, it will display larger values of $\alpha$. Dust populations dominated by large dust particles, on the contrary, have much lower values of $\beta$ \citep[e.g][]{dal01,bir18}, which results in a lower $\alpha$. Therefore, the spectral index profile of HD 169142 could be explained if small dust particles are filling the gaps, while large dust grains are accumulating in the ring substructures and are dominating their (sub)mm emission. However, (sub)mm dust emission can also present low values of $\alpha$ if its optical depth ($\tau$) is close to (or larger than) unity. Both effects, low $\beta$ and high $\tau$, are in principle degenerate and indistinguishable with just two wavelengths. Nevertheless, by analyzing the data at 0.89 mm, 1.3 mm, and 3.1 mm, we should be able to pose strict limits on the radial distribution of $\tau$ and $\beta$. We note that the observations at these three bands have different uv coverages and spatial resolutions, which could affect the results if the analysis was performed in the image domain. In order to avoid these issues, while also extracting as much information as possible from the different spatial scales probed by the interferometer, we perform our analysis in the visibility domain.

\section{Visibility Modeling}\label{sec:modeling}

In order to analyze our multi-wavelength observations of HD 169142, we aim at fitting an analytical model of the radial intensity profile of the disk to the real part of the observed deprojected visibilities. Following studies of other protoplanetary disks \citep[e.g.][]{zha16,pin18,mac18}, we assume that the disk emission is axisymmetric, so that the visibility profile of the disk can be computed as the Hankel transform of its radial intensity profile \citep{pea99}. We can then compare the model and observed visibilities, and explore the parameter space following a Bayesian approach with a Markov Chain Monte Carlo (MCMC) algorithm. We note that the small azimuthal asymmetries in the inner ring (B1) are just marginally resolved, so we do not expect them to have important effects at the spatial scales traced by our observations. Therefore, an axisymmetric disk is a good approximation and should give us a good estimate of the average radial structure of the disk. 

Instead of fitting a surface brightness profile at each band separately, we fit all the observed visibilities at the same time. To do this, we assume that the disk is geometrically thin and vertically isothermal -- a good approximation at (sub)mm wavelengths, since the emission is dominated by the large dust grains that are settled onto a thin layer in the disk mid-plane --, and that the intensity at each radius $r$ can be calculated as:
\begin{equation}
   I_{\nu} (r) = B_{\nu}(T_d(r)) \, \,(1 - e^{-\tau_{\nu} (r)}), \\
\end{equation}
\begin{equation}
   \tau_{\nu} (r) = \tau_{0} (r) \left(\frac{\nu}{\nu_0}\right)^{\beta(r)}, \\
\end{equation} 
where $B_{\nu}$ is the Planck function, $T_d$ is the dust temperature, $\tau_0$ is a reference optical depth at $\nu_0=345$ GHz, and $\beta$ is the power of the dust opacity law ($\kappa_{\nu}\propto\nu^{\beta}$). In principle, the three free parameters at each radius are therefore $T_d$, $\tau_0$, and $\beta$, for which we will presume a certain functionality. We note that we have made the common assumption that the main source of opacity is the absorption opacity, without considering the scattering component. For a discussion of the possible effects of the latter component, see \citet{sie19}.

Firstly, we assume that the dust temperature follows a power-law profile:
\begin{equation}
   T_{d} (r) = T_0 \left(\frac{r}{r_0}\right)^{q}, \\
\end{equation}
where $T_0$ is the dust temperature at the reference radius $r_0 = 10$ au. In a flared disk in radiative equilibrium, the mid-plane temperature profile is expected to follow a power-law $T\propto r^{-0.5}$. Instead of fixing $q$ to $-0.5$, we let this parameter vary in our model to account for possible deviations from this simplified view.

For $\tau_0$ and $\beta$, we aim at choosing functions that can be flexible enough to reproduce our observations without overfitting them. We do this by assuming that the disk structure is composed by a \textit{base} extended component and a set of rings where higher densities and/or smaller or larger dust particles may be concentrated. The \textit{base} component is modeled as power-laws for both $\tau_0$ and $\beta$:
\begin{equation}
   \tau_{0}^{base} (r) = a^{(base)} \left(\frac{r}{r_0}\right)^{s}, \\
\end{equation}
\begin{equation}
   \beta^{base} (r) = b^{(0)} + b^{(base)}\left(\frac{r}{r_0}\right)^{t}, \\
\end{equation}
where we take $r_0=10$ au. We note that we include an extra parameter ($b^{(0)}$) for $\beta^{base}$ since $\beta$ is not expected to reach 0, as opposed to $\tau_{0}$. This results in 5 free parameters for the \textit{base} component. For the  rings, we use radially asymmetric Gaussians (i.e., Gaussian rings with independent inner and outer widths):
\begin{equation}
   \tau_{0}^{(i)} (r) = \begin{dcases}
   a^{(i)}~\textrm{exp} \left[ -\left( \frac{r - x^{(i)}}{\sqrt{2}~\sigma_{-}^{(\tau, i)}}\right)^2 \right], & r \leqslant x^{(i)} \\
   a^{(i)}~\textrm{exp} \left[ -\left( \frac{r - x^{(i)}}{\sqrt{2}~\sigma_{+}^{(\tau, i)}}\right)^2 \right], & r > x^{(i)} \\
\end{dcases}
\end{equation}
\begin{equation}
   \beta^{(i)} (r) = \begin{dcases}
   b^{(i)}~\textrm{exp} \left[ -\left( \frac{r - x^{(i)}}{\sqrt{2}~\sigma_{-}^{(\beta, i)}}\right)^2 \right], & r \leqslant x^{(i)} \\
   b^{(i)}~\textrm{exp} \left[ -\left( \frac{r - x^{(i)}}{\sqrt{2}~\sigma_{+}^{(\beta, i)}}\right)^2 \right], & r > x^{(i)} \\
\end{dcases}
\end{equation}

where $i$ denotes each ring; $\sigma_{-}$ and $\sigma_{+}$ are the inner and outer widths, respectively, which can be different for $\tau_{0}$ and $\beta$ (denoted by their superindex); $a$ and $b$ are the peak of the Gaussians for $\tau_{0}$ and $\beta$, respectively; and $x$ is the radius of the ring. We force the rings in $\tau_{0}^{(i)}$ and $\beta^{(i)}$ to be centered at the same position $x^{(i)}$, but we let the other parameters (scale, inner width, and outer width) vary as free parameters.
This results in 3 free parameters per ring for $\tau_{0}$, 3 per ring for $\beta$, plus 1 parameter per ring for the position of its peak, for a total of 7 free parameters per ring. Based on the shape of the radial profiles obtained from the cleaned images (Fig. \ref{fig:profiles}), we use three rings in our model to reproduce B1, B2/3, and B4. 
Motivated by the recent high spatial resolution ALMA observations that resolved the outer disk into three separate rings \citep{per19}, we ran an additional model using four Gaussian rings. However, due to the lower spatial resolution of our data, our model with four rings does not reproduce the triple ring morphology of the outer disk, instead predicting a similar morphology to the case with three rings. Therefore, in the following we use the results of our model with three Gaussian rings. 

Finally, we add the \textit{base} and ring components of $\tau_{0}$ and $\beta$ and include them in eq. (2), which is then used in eq. (1) together with the temperature power-law (eq. (3)) to predict the intensity radial profile at each frequency. The combination of the two type of components of our model (power-law plus three Gaussian rings) is flexible enough to reproduce the observed emission while exploring different origins for the rings. If the rings are not associated with spatial variations of the dust size distribution, we expect the model to reproduce them with substantial increases in $\tau_0$ (i.e., $a^{(i)}>>0$), but no changes in $\beta$ (i.e., the radial profile in $\beta$ being dominated by the \textit{base} component, with $b^{(i)}\sim0$). On the contrary, if the rings are in fact produced by gas pressure bumps that are trapping and accumulating large dust grains, we expect to have increases in $\tau_0$ coupled with decreases in $\beta$ at the rings (i.e., $b^{(i)}<<0$). We note that by using $\tau_0$ and $\beta$ we avoid making any initial assumptions about the dust composition, which would have been necessary if we aimed at directly using the dust opacity ($\kappa_{\nu}$) in our model.

In addition to the radial profiles of $T_d$, $\tau_0$, and $\beta$ --which determine the radial intensity profile--, we include in our model the orientation of the disk (inclination, $i$, and position angle of major axis, PA), and potential offsets from the expected position of the star for each observation ($\Delta$RA and $\Delta$DEC). Finally, we take into account the potential effects of the systematic flux calibration uncertainty by including the systematic errors in our model. These systematics are, in principle, unconstrainable, but by including them in our MCMC as nuisance parameters, we can explore their effects and include the systematic uncertainty in the posterior of the rest of parameters.

In summary, our model (with three rings) has 51 free parameters (28 parameters for the $T_d$, $\tau_0$, and $\beta$ profiles; 23 for disk orientation, phase center shifts, and systematic flux corrections). We separately fit the upper sideband and lower sideband of the band 7 and band 3 observations (the band 6 data only had continuum spectral windows in one of the sidebands), so we end up fitting data at 5 different frequencies. In order to reduce the computational time of the model and ensure that each observation has a similar weight during the model fitting, we bin the deprojected visibilities into 2000 bins.
We use the MCMC ensemble samplers with affine invariance \citep{goo10} in the \texttt{EMCEE} package \citep{for13}. We set Gaussian priors for the systematic errors in the flux calibration based on the nominal errors for ALMA at each band (i.e., $10\%$ at Band 7, and $5\%$ at Band 6 and 3). We also use Gaussian priors for the position offsets based on the expected astrometry uncertainty for ALMA ($\sim5\%$ of the resolution), and for the inclination and PA of the disk using previous estimates from line observations ($i=13\pm1^{\circ}$ and PA$=5\pm5^{\circ}$; \citealp{ram06}). For the rest of the parameters we set flat priors between reasonable values. For the width of the Gaussian rings, we chose the lower limit of the flat priors to be $\sim0\rlap.''01$ ($\sim1/10$ of the beam size), since our observations should not be able to probe smaller spatial scales. This resulted in some chains in the model being limited by the priors, indicating that those structures are partially unresolved. Finally, in order to avoid obtaining non-physical values of $\beta$, we set a flat prior that limits $\beta$ to be between 0.2 and 3.1 at all radii. These limits are based on the values of $\beta$ obtained using the dust opacities from \citet{bir18}, and exploring different power-law size distributions with its power ranging from 1.5 to 4.5, and the maximum grain size from 0.1 $\mu$m to 1 m. We used 200 walkers and ran 100,000 iterations, which was enough to ensure that, despite the relatively large number of free parameters, the chains of the MCMC had converged.

\subsection{Model Results}\label{subsec:modelresults}

\begin{figure*}
\centering
\figurenum{4}
\includegraphics[width=\textwidth]{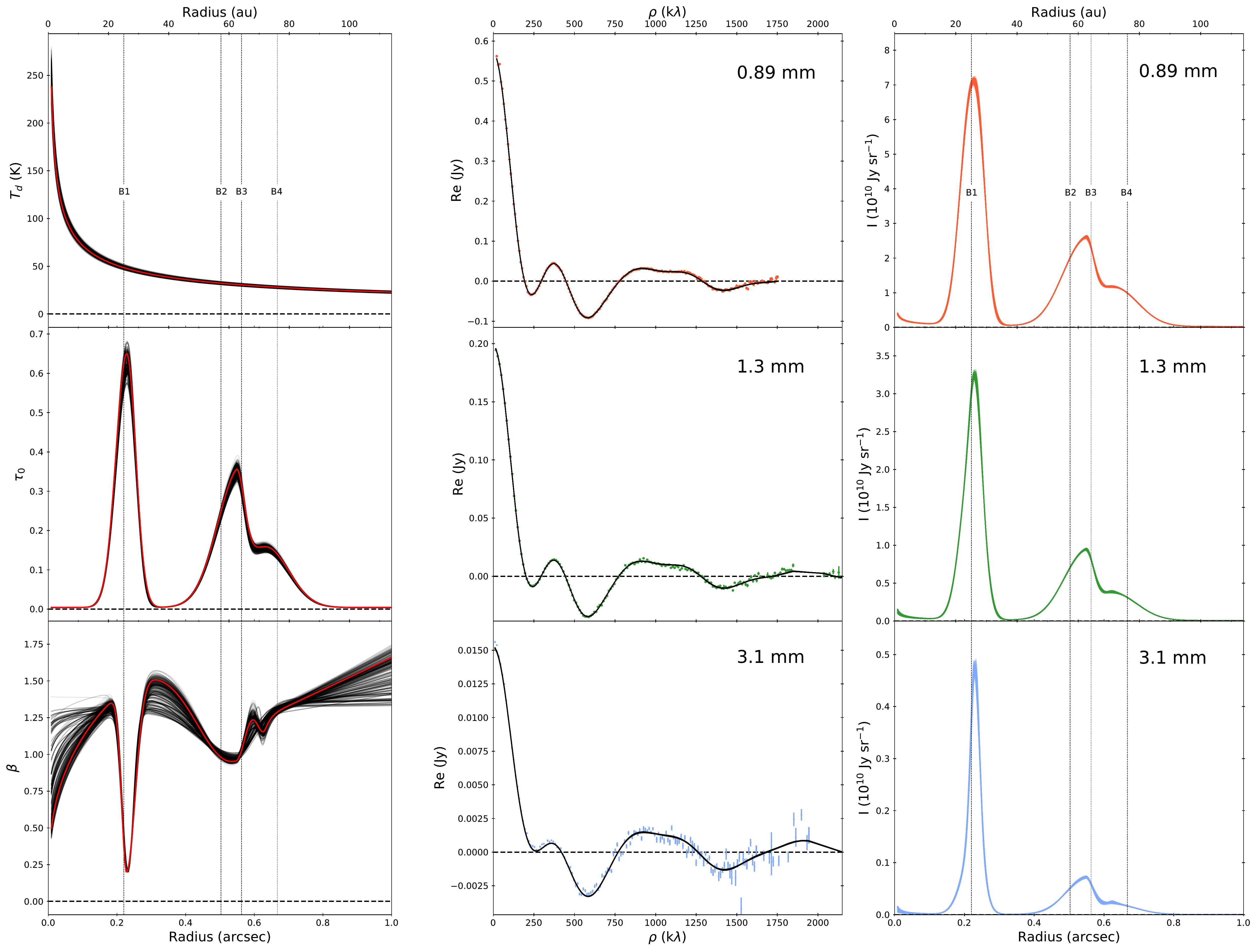}
\label{fig:mcmc}
\caption{Results from our analytical model fitting of the observed visibilities at 0.89 mm, 1.3 mm, and 3.1 mm. \textit{Left:} Radial profiles of dust temperature ($T_d$; top), optical depth at 345 GHz ($\tau_0$; middle), and $\beta$ (bottom) obtained by our model. The red solid line indicates the maximum likelihood model, while the black lines represent 500 random chains chosen from the MCMC posteriors. The dashed vertical lines show the radius of the 4 rings reported by \citet{per19}. \textit{Middle:} Real part of the deprojected visibilities at 0.89 mm (top), 1.3 mm (middle), and 3.1 mm (bottom). The black lines show the predicted visibilities for the 500 random chains. The 0.89 mm and 3.1 mm panels only show the USB and LSB of the observations, respectively. Additionally, the 1.3 mm panel displays only one of the observed epochs used in the modeling. The other epochs and sidebands are also fitted but not shown, since they would appear at different scales because of the difference in frequency and/or the different systematic flux calibration correction. \textit{Right:} Intensity profiles at the same three wavelengths. In each panel we plot 500 profiles obtained from the 500 random chains mentioned above. As in the left panels, the dashed vertical lines show the central positions of the 4 rings in the disk. }
\end{figure*}

\begin{figure*}
\centering
\figurenum{5}
\includegraphics[width=\textwidth]{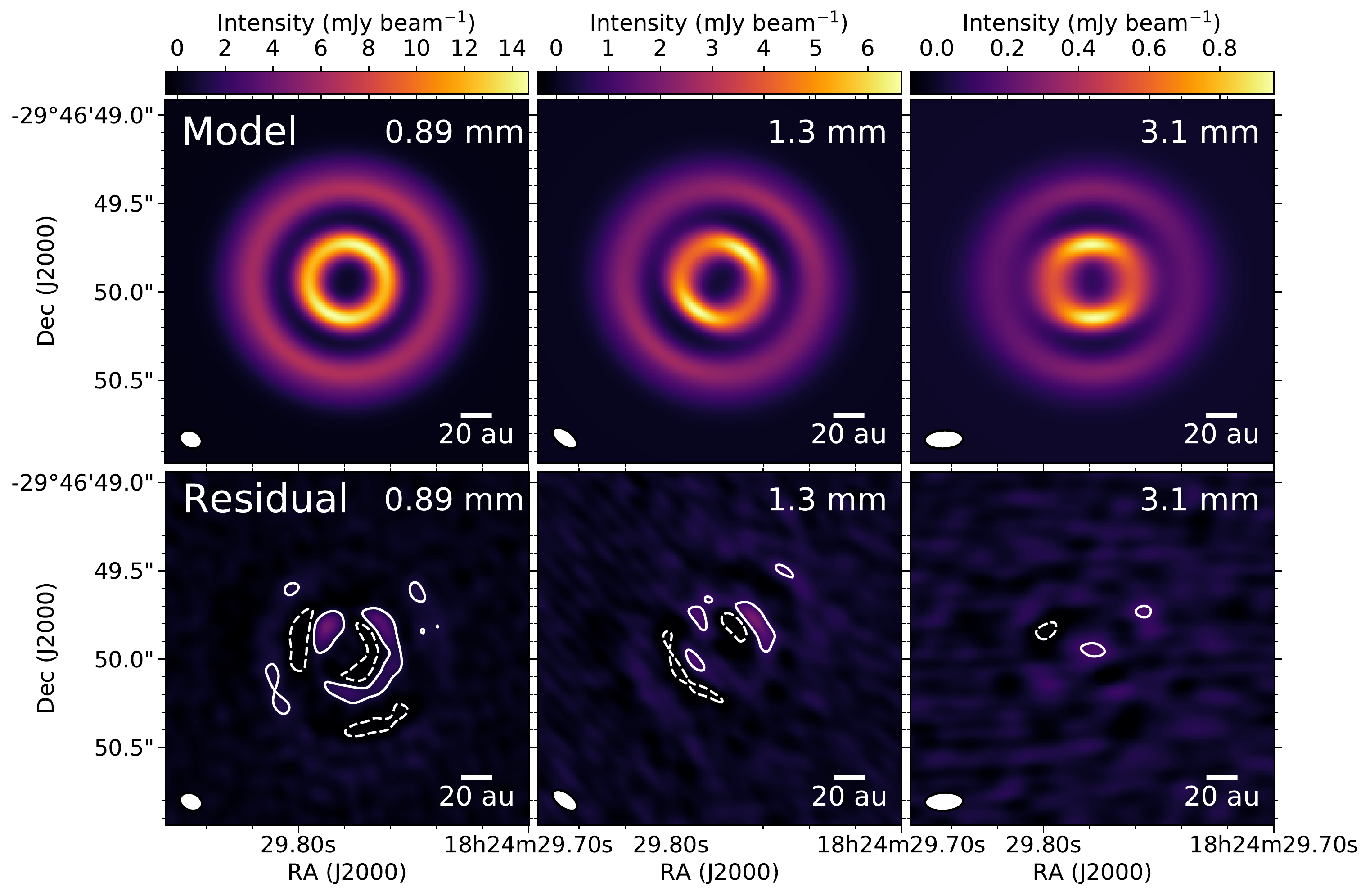}
\label{fig:residuals}
\caption{Images of the results from our analytical model fitting of the observed visibilities at 0.89 mm, 1.3 mm, and 3.1 mm (from left to right). The top panels show the model images convolved to the beam of the observations (see Fig. \ref{fig:maps}). The bottom panels display the residuals (observation - model). The white contours in the bottom panels show the $5\sigma$ (solid) and $-5\sigma$ (dashed) levels at each band.
}
\end{figure*}

The radial profiles of $\tau_0$, $\beta$, and $T_d$ obtained from our model are shown in Fig. \ref{fig:mcmc}, together with the real part of the deprojected visibilities and the predicted intensity profiles at 0.89 mm, 1.3 mm, and 3.1 mm. The median and percentiles of the model parameters of these profiles are listed in Table \ref{Tab:model}. Images of our model and the residuals at these three wavelengths are shown in Fig. \ref{fig:residuals}. As can be seen, our model successfully reproduces the observations in great detail, with the residuals only appearing mostly due to the small departures from axisymmetry of the inner ring.

\floattable
\begin{deluxetable}{lccc}
\tablecaption{Model parameters.  \label{Tab:model}}
\tablehead{\colhead{Parameter} & \multicolumn{3}{c}{ Median, $16\%$, and $84\%$ percentiles} }
\startdata
\multicolumn{4}{c}{Temperature profile}\\
\hline
$T_0$ (K) & \multicolumn{3}{c}{ $77^{+2}_{-2}$ } \\
$q$ & \multicolumn{3}{c}{ $-0.501^{+0.011}_{-0.011}$ } \\
\hline
\hline
\multicolumn{4}{c}{Radial profile of $\tau_0$ }\\
\hline
\multicolumn{4}{c}{\textit{base} component} \\
\hline
$a^{(base)}$ & \multicolumn{3}{c}{ $0.00424^{+0.00016}_{-0.00020}$ } \\
$s$ & \multicolumn{3}{c}{ $-0.003^{+0.003}_{-0.006}$ } \\
\hline
& Ring 1 & Ring 2 & Ring 3 \\
\hline
$a^{(i)}$  & $0.63^{+0.02}_{-0.02}$  & $0.271^{+0.016}_{-0.014}$ & $0.153^{+0.006}_{-0.007}$ \\
$\sigma^{(\tau, i)}_{-}$  &  $0\rlap.''0317^{+0\rlap.''0010}_{-0\rlap.''0005}$  & $0\rlap.''0619^{+0\rlap.''006}_{-0\rlap.''008}$ & $0\rlap.''070^{+0\rlap.''0007}_{-0\rlap.''0005}$ \\
$\sigma^{(\tau, i)}_{+ }$ & $0\rlap.''0258^{+0\rlap.''0005}_{-0\rlap.''0017}$ & $0\rlap.''0218^{+0\rlap.''0014}_{-0\rlap.''0017}$ & $0\rlap.''067^{+0\rlap.''0011}_{-0\rlap.''0014}$ \\
$x^{(i)}$ & $0\rlap.''2288^{+0\rlap.''0021}_{-0\rlap.''0007}$ & $0\rlap.''5474^{+0\rlap.''0014}_{-0\rlap.''0009}$ & $0\rlap.''631^{+0\rlap.''003}_{-0\rlap.''002}$ \\
\hline
\hline
\multicolumn{4}{c}{ Radial profile of $\beta$\tablenotemark{$\dagger$}}\\
\hline
\multicolumn{4}{c}{\textit{base} component}\\
\hline
$b^{(0)}$ & \multicolumn{3}{c}{ $0.06^{+0.30}_{-0.05}$ } \\
$b^{(base)}$ & \multicolumn{3}{c}{ $1.00^{+0.07}_{-0.2}$ } \\
$t$ & \multicolumn{3}{c}{ $0.32^{+0.04}_{-0.12}$ } \\
\hline
& Ring 1 & Ring 2 & Ring 3 \\
\hline
 $b^{(i)}$ &  $-1.22^{+0.06}_{-0.04}$ & $-0.92^{+0.33}_{-0.10}$ & $-0.70^{+0.43}_{-0.13}$ \\
$\sigma^{(\beta, i)}_{-}$ & $0\rlap.''0151^{+0\rlap.''0013}_{-0\rlap.''0011}$ & $0\rlap.''113^{+0\rlap.''005}_{-0\rlap.''006}$ & $0\rlap.''025^{+0\rlap.''004}_{-0\rlap.''004}$ \\
$\sigma^{(\beta, i)}_{+}$ & $0\rlap.''0235^{+0\rlap.''0014}_{-0\rlap.''0042}$ & $0\rlap.''040^{+0\rlap.''006}_{-0\rlap.''004}$ & $0\rlap.''95^{+0\rlap.''06}_{-0\rlap.''14}$ \\
\enddata
\tablenotetext{\dagger}{The rings in the $\beta$ profile have the same radius as the rings in the $\tau_0$ profile.}
\end{deluxetable}

We find systematic flux calibration corrections that are within the expected uncertainties. For the phase center shifts we obtain values $\lesssim \pm20$ mas, also consistent with the expected astrometric uncertainties of ALMA. The inclination and PA of the model with maximum likelihood are $\sim12.3^{\circ}$ and $\sim4.8^{\circ}$, respectively, consistent with the values estimated from line observations ($13^{\circ}\pm1^{\circ}$ and $5^{\circ}\pm5^{\circ}$; \citealp{ram06}).

Our model predicts a temperature profile with a reference temperature at 10 au of $77^{+2}_{-2}$ K (median, 16th, and 84th percentiles) and a power $-0.501^{+0.011}_{-0.011}$. This profile is consistent with the expected slope for a flared irradiated disk, indicating that the temperature profile of the disk is not significantly affected by the presence of the deep gaps in the disk. We can further compare our temperature profile to the expected profile for such a disk \citep{chi97,dul01}:
\begin{equation}
T (r) = \left( \frac{\varphi L_{\star}}{8\pi r^2 \sigma_{SB}} \right)^{0.25} , \\
\end{equation}
where $\varphi$ is the flaring angle of the disk (as defined in \citealp{chi97}), $L_{\star}$ is the stellar luminosity, and $\sigma_{SB}$ is the Steffan-Boltzmann constant. Assuming $L_{\star}=10~L_{\odot}$ \citep{fed17}, our temperature profile would imply $\varphi=0.031$, or a height of the disk photosphere $H_{ph}\sim0.11r$ at $r=100$ au, consistent with the standard range of values found for Class II objects.

The \textit{base} component of $\tau_0$ has almost no contribution, and as a consequence the inner cavity (D1) and the gap (D2) are almost completely devoid of emission. Furthermore, the general trend of $\beta$ is to increase with radius (see Fig. \ref{fig:mcmc}), consistently with the expected effects of radial migration. The values of $\beta$ at the innermost radii tend to decrease, probably as a consequence of the emission from the inner disk, which was recently detected at mm wavelengths in high resolution observations at $r\leqslant1$ au \citep{per19}. In order to properly account for this component our model would need to include an extra Gaussian centered at $r=0$ au, but our observations would not have the necessary resolution to constrain this component. This is in fact shown by the high uncertainty in $\beta$ at $r\lesssim10$ au. Therefore, we note that the exact behavior of $\beta$ in this region should be taken with caution.

On the other hand, our model finds that the multiple rings in HD 169142 can only be explained with increases in $\tau$ coupled with substantial decreases in $\beta$.
The inner ring is centered at $\sim0\rlap.''2287^{+0\rlap.''0021}_{-0\rlap.''0007}$ and displays a sharp and narrow decrease in $\beta$ down to values $\sim0.2$. This low $\beta$ is coupled with an increase in $\tau_{0}$. The ring is wider in the $\tau_0$ profile than in the $\beta$ one (see Fig. \ref{fig:mcmc}). As a result, the predicted emission from the inner ring is significantly narrower at longer wavelengths. 
For the outer disk, the two rings show increases in $\tau_0$ coupled with decreases in $\beta$ that are not as deep as in the inner ring. Our model predicts its second ring (B2/3) at $\sim0\rlap.''5474^{+0\rlap.''0014}_{-0\rlap.''0009}$, while the outermost ring (B4) is centered at $\sim0\rlap.''631^{+0\rlap.''003}_{-0\rlap.''002}$. The latter only shows a very slight decrease in $\beta$, which overall indicates that strength of the accumulations of large particles in HD 169142 decreases with radius.

We note that given the finite resolution of our observations, our model results on $\tau$ and $\beta$ are smeared out. This effect is minimized by fitting the model in the visibility domain (see \S\ref{subsec:optdepth}), but our resolution still limits the spatial scales of the substructures that we are able to characterize. In particular, the angular resolution of our data ($0\rlap.''1$ and $0\rlap.''2$) is larger than the pressure scale height ($\sim0\rlap.''03$ at 50 au, from our temperature profile), which is the expected size of disk substructures associated with gas perturbations \citep[e.g.][]{don17}.

\subsubsection{Dust Surface Density and Particle Size Distribution}

We note that our multi-wavelength analysis is based on $\tau_0$ and $\beta$ rather than on the dust surface density ($\Sigma_d$) and the dust opacity ($\kappa_{\nu}$). In this way, we obtain a more direct and model independent estimate of the disk properties. However, an increase of $\tau_0$ in our model could be the result of higher $\Sigma_d$ and/or higher $\kappa_0$. At the same time, lower values of $\beta$ can be obtained with larger maximum grain sizes, or with flatter dust particle size distributions \citep{dal01,bir18}. 

\begin{figure*}[t]
\centering
\figurenum{6}
\includegraphics[width=\textwidth]{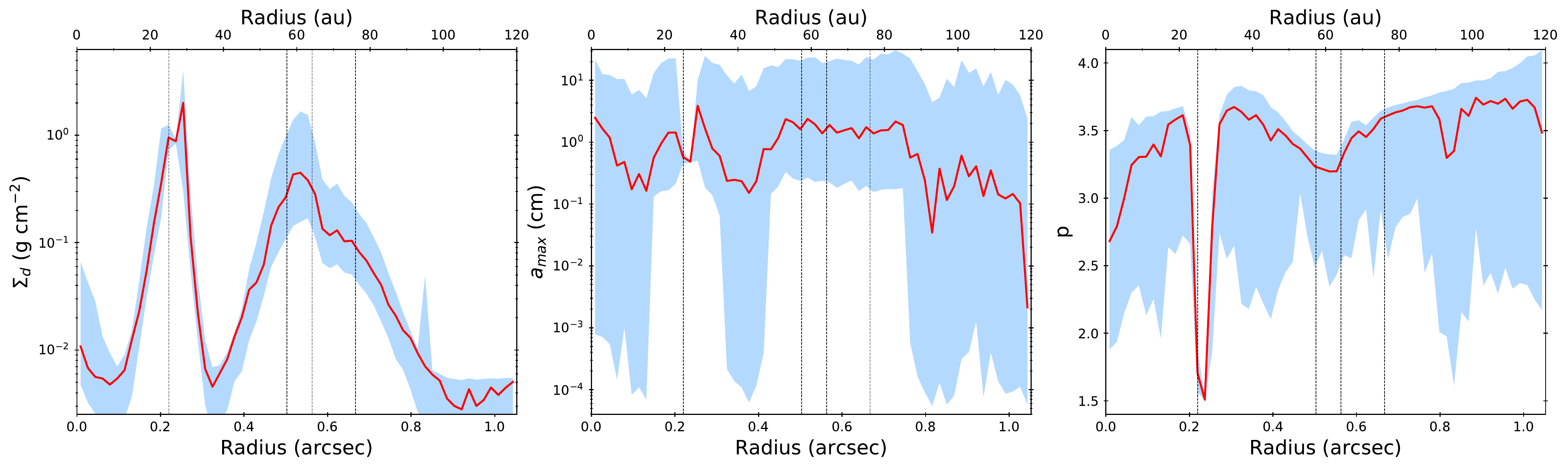}
\label{fig:dust}
\caption{Radial profiles of dust surface density ($\Sigma_d$, left), maximum grain size ($a_{max}$, middle), and power of particle size distribution ($p$, right) for the disk of HD 169142. These profiles are obtained by exploring the range of parameters that can explain the $\tau_0$ and $\beta$ obtained by our model. The red solid line indicates the median of the posteriors at each radius, while the blue colored shade shows the region between the 16th and 84th percentiles. The dashed vertical lines show the radius of the 4 rings reported by \citet{per19}.  }
\end{figure*}

We can further explore the implications of our results in terms of the dust size distribution by assuming a particular dust composition. We use the dust composition employed in the DSHARP ALMA Large Program \citep{bir18} and assume the typical power-law for the particle size distribution $n(a)\propto a^{-p}$. Using the absorption opacities and Python tools provided by \citet{bir18}\footnote{Opacities and Python scripts are availabe at \href{https://github.com/birnstiel/dsharp_opac}{https://github.com/birnstiel/dsharp\_opac}} we estimate $\kappa_0$ and $\beta_{0.89-3~{\rm mm}}$ while varying the maximum grain size ($a_{max}$) and power of the size distribution ($p$). Thereafter we can try to reproduce our predicted profiles of $\tau_0$ and $\beta$, using $\Sigma_d$, $a_{max}$, and $p$ as the three free parameters at each radius. This approach results in a model that is in principle unconstrained, with three free parameters for two data points per radius, but that can be used to discard some combinations of parameters (i.e., low $\beta$ and high $\tau_0$ cannot be obtained with micron-sized particles). We explore the parameter space using an MCMC. For each radius we run an MCMC with 50 walkers for 10000 iterations, with a burn-in phase of 8000 iterations. The median, 16th, and 84th percentiles of the posteriors at each radius are shown in Figure \ref{fig:dust}. Overall, $\Sigma_d$ appears to be most sensitive to $\tau_0$, $p$ to $\beta$, and $a_{max}$ to both $\tau_0$ and $\beta$. 

Despite the high uncertainties, we are able to set some constraints on the three parameters, especially at the position of the inner ring and of the outer disk. 
In D1, D2, and at  $r>100$ au, $\Sigma_d$ is the parameter that is more tightly constrained, showing values close to 0. In these same regions of the disk, $p$ is consistent with the ISM value of 3.5, although lower values would also be possible. On the other hand, $a_{max}$ remains completely unconstrained at these radii.
Close to $r=0$ au our analysis predicts a possible slight increase in $\Sigma_d$ and a slight decrease in $p$. These features are associated with the contribution from the inner disk that is likely not well constrained (see \S\ref{subsec:modelresults}), so we advise caution when interpreting the results in these region.

The extremely low values of $\beta$ found at the peak of B1 can only be reproduced with a narrow range of values in the parameter space: $p=1.509^{+0.014}_{-0.006}$, $a_{max}=4.85^{+0.15}_{-0.15}$ mm, and $\Sigma_d \sim0.89^{+0.04}_{-0.04}$ g cm$^{-2}$. In particular, the low value of $p$ clearly points toward the presence of a remarkably strong accumulation of large pebbles. In the outer disk, $\Sigma_d$ shows the double peak morphology displayed by B2/3 and B4 in the $\tau_0$ profile, with the two peaks in the range $0.16-1.7$ g cm$^{-2}$ and $0.063-0.35$ g cm$^{-2}$. At these same radii, $a_{max}$ displays a plateau-like shape around 1 cm, with possible values ranging from $\sim2$ mm to $\sim20$ cm, and without showing any significant changes between B2/3 and B4. At the position of B2/3, $p$ decreases slightly down to $\sim3.2$, but it appears to go up to $\sim3.5$ in B4. Overall, this suggests that the rings B2/3 and B4 are produced by accumulations of large dust grains, but at a much lesser degree than in B1. It is important to note that B2/3 is in fact composed of two unresolved rings \citep{per19}, so it is possible that our estimates of $\beta$ are underestimated and that narrower and stronger accumulations of large particles are present at B2 and B3 separately. 

Finally, by integrating the dust surface density profile over the disk area we estimate a total dust mass of $160^{+250}_{-90}$ M$_{\bigoplus}$. Despite the underlying uncertainty introduced by assuming a certain dust composition, our dust mass estimate is significantly accurate, since this method accounts not only for radial changes in the optical depth, but also for possible variations in dust opacity that are not usually considered.
Based on modeling of ALMA $^{12}$CO, $^{13}$CO, and C$^{18}$O line emission, \citet{fed17} estimated a gas mass of $19$ M$_J$. Therefore, our estimated dust mass would imply a dust-to-gas mass ratio of $0.03^{+0.04}_{-0.02}$, which is consistent with the ISM value of 0.01, but suggests that it could be substantially higher. A dust-to-gas mass ratio higher than 0.01 would be reasonable for a $\sim10$ Myr old disk such as HD 169142, since central photoevaporation should have had plenty of time to disperse a considerable amount of gas from the disk \citep{owe13}. 
Nevertheless, we note that this high dust-to-gas mass ratio does not include an uncertainty in the gas mass, and the depletion of CO in the disk could result in a significant underestimate of the total gas mass \citep[e.g.,][]{mio17}.

\section{Discussion}\label{sec:discussion}

\subsection{Optical Depth}\label{subsec:optdepth}

\begin{figure*}
\centering
\figurenum{7}
\includegraphics[width=\textwidth]{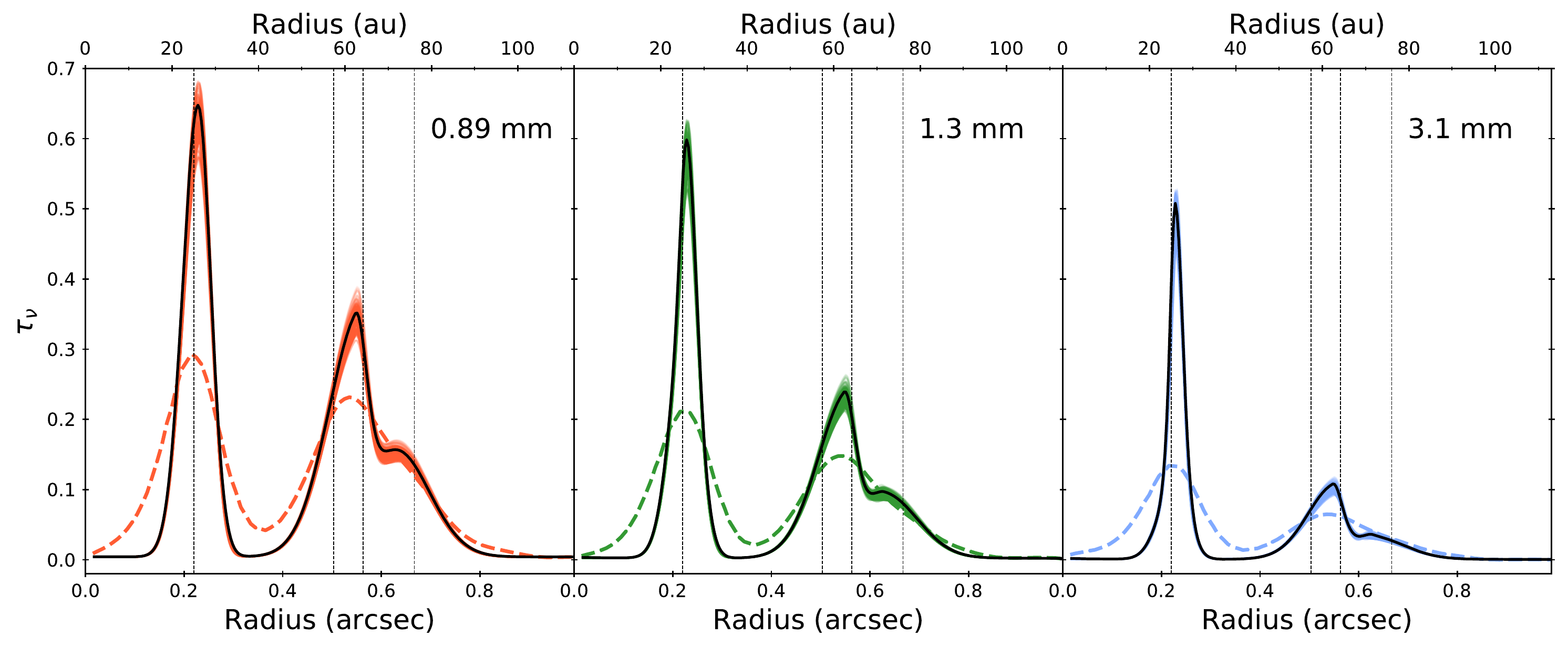}
\label{fig:taus}
\caption{Radial profiles of optical depth at 0.89 mm (left), 1.3 mm (middle), and 3.1 mm (right). The colored solid lines in each panel show the optical depth of one chain of the MCMC modeling. The black solid lines indicate the optical depths for the maximum likelihood chain in our MCMC. The dashed lines represent the optical depths profile obtained from the cleaned images. The dashed vertical lines show the radius of the 4 rings reported by \citet{per19}. These are computed from our images in Fig.\ref{fig:maps} after convolving them to the beam of our 3 mm maps, estimating the radial intensity profiles along the minor axis of the beam, and assuming the temperature profile obtained from the maximum likelihood chain of our MCMC. }
\end{figure*}

In order to reproduce the multi-wavelength observations of HD 169142, our model predicts a radial profile of $\beta$ and $\tau$ at 345 GHz, with which we can obtain the predicted $\tau$ at the frequencies of our observations. These optical depth profiles are shown in Figure \ref{fig:taus}. 

We find optical depths that are lower than unity at all radii and bands, but they show values $>0.1$ in the inner ring and outer disk even at 3 mm. In particular, the optical depth in B1 is close to 1 at 0.89 mm and 1.3 mm. This result is consistent with the optical depths found in ring substructures of other protoplanetary disks \citep[e.g.][]{car16,hua18a,dul18} and it has one important implication: the usual assumption that the (sub)mm emission of protoplanetary disks is optically thin could be significantly inaccurate. This assumption has also been called into question in other recent studies \citep[e.g.][] {tri17,and18b}, and it could have important effects. The dust emission is usually assumed to be optically thin in order to obtain dust masses from (sub)mm surveys. At the same time, the (sub)mm spectral index of disks has been used to estimate $\beta$ and hence the level of grain growth in disks. If a significant fraction of the disk is optically thick, these mass estimates could be substantially underestimated, and apparent signs of grain growth (i.e., $\alpha\lesssim2.5$) could be incorrectly identified in massive disks (see also \citealp{ric12}). 

We note that even when resolved multi-wavelength observations are available, angular resolution can play an important role in determining accurate optical depths. If the optically thick regions of the disk are compact -- as expected for dust traps in radial pressure bumps (see \S\ref{subsec:traps}) -- observations without sufficient spatial resolution would smear the emission, resulting in apparent lower optical depths. 

In our case, we minimize this effect by directly modeling the observed visibilities instead of the cleaned images. In this way, we use the information of all the spatial scales probed by our observations. Figure \ref{fig:taus} shows a comparison of the optical depth profiles of our models (solid lines) and the profiles obtained from the cleaned images (dashed lines). As shown in the figure, using the radial profiles from the cleaned images would have resulted in a significant underestimation of the peak optical depths by a factor $\sim3$. For this same reason, the peak optical depths estimated by our model are in principle lower limits, and even larger optical depths could be present at smaller spatial scales than the ones probed by our observations and modeling.

\subsection{Dust Grain Size Distribution}\label{subsec:traps}

As mentioned in \S\ref{sec:results}, the spectral index profile of the disk of HD 169142 displays higher values in the gaps than in the inner ring and outer disk. This behavior could in principle be explained with two different scenarios: the accumulation of larger grains in the rings while smaller grains fill the gaps, and/or optical depths close to 1 in the rings. Previous studies have found similar trends in the spectral index of other disks \citep[e.g.][]{tsu16,hua18a}, but so far very few studies have successfully found unambiguous evidence of accumulations of large dust particles in ring substructures \citep[e.g.][]{car16,liu17,mac18}. By analyzing our multi-wavelength observations we are able to disentangle both effects and demonstrate that the inner ring and the outer disk of HD 169142 are produced by an increase in $\tau$ coupled with a decrease in $\beta$. This behavior indicates that large dust particles (indicated by a low $\beta$) are being accumulated in the ring substructures, confirming previous results from the modeling of mm observations at individual wavelengths \citep{oso14,fed17}.

Furthermore, our estimate of the dust surface density and particle size distribution shows low $\Sigma_d$ and ISM-like values of $p$ in the gaps (D1 and D2) and at $r>100$ au. These trends are consistent with these regions being almost completely depleted of large dust particles, as also indicated by recent high resolution 1.3 mm observations \citep{per19}.  However, D1 and D2 still harbor some amount of micron-sized dust particles, since recent polarimetric observations show that the gaps are shallower at near-IR wavelengths \citep{mon17,poh17,ber18}. The ISM-like values of $p$ that we find are consistent with this scenario, but since our observations are not sensitive to micron-sized particles, we are unable to constrain the amount of small particles filling the gaps.

On the other hand, our analysis of the particle size distribution also indicates that the low values of $\beta$, coupled with high values of $\tau$, at the inner ring and outer disk can be reproduced with dust populations that have a flatter size distribution than the ISM ($p<3.5$), and have maximum grain sizes between 2 mm and 20 cm. 
These results strongly indicate that the ring substructures in HD 169142 are the result of accumulations of large dust grains. Such buildups of solids might be associated with increases in growth efficiency beyond certain volatile snowlines, but the more likely scenarios involve the presence of gas pressure bumps that are able to trap the large dust particles (see \S\ref{subsec:origin}). These dust traps have been hypothesized to be an important ingredient in the planet formation process, since they can stop the radial drift of large particles, concentrate them, and enable them to grow up to planetesimal sizes \citep{pin12}.
The discovery of the likely ubiquity of disk substructures has represented a strong support to this theory \citep{and18a,dul18,lon18,van19}, but few studies have been able to find unequivocal evidence of their role as dust traps (e.g., \citealp{liu17,mac18}, this work). If similar evidence is found for a larger sample of objects, it could represent the final confirmation that ring substructures represent the solution to the drift and fragmentation barriers in the planetary formation process.

In the case of HD 169142, the extremely low $p$ at B1 (see Fig. \ref{fig:dust}) is particularly interesting. Such a flat size distribution could only be formed in an extremely tight accumulation of particles that has significantly enhanced the dust growth efficiency, as fragmentation is expected to move $p$ toward $\sim3.5$ \citep{bir11}. According to our results, the dust surface density at this position reaches $\sim1$ g cm$^{-2}$, which would imply dust-to-gas mass ratios $\sim1$ when compared with the gas surface density estimated by \citet{fed17}. If confirmed, this inner ring would represent an ideal location to trigger the streaming instability and hence form the seeds for new young planets \citep{you05,auf18}. It is in fact possible that the streaming instability was triggered in the past at this ring, thus resulting in a \textit{run-away} growth process responsible for the low values of $p$. Interestingly, the predicted $a_{max}$ at this position is not particularly high ($\sim5$ mm), despite the low value of $p$ \citep{abo18}. Nevertheless, we have assumed a certain dust composition, which could affect the results. In addition, our assumption that the dust size distribution can be described by a power-law with a single power might not be appropriate for such accumulations of particles. Lastly, the streaming instability might have induced the formation of small azimuthal clumps in the ring that are not resolved in our data, and where larger particles might be accumulating. Evidence of these possible asymmetries has in fact been revealed by the VLA \citep{mac17} and ALMA \citep{per19}. A multi-wavelength analysis at higher spatial resolution will be needed to confirm the flat particle size distribution and remarkably strong dust traps at the inner ring of HD 169142.

Finally, we note that our results are derived purely from the dust continuum emission, without including information about the gas. A more complete description of the dust trapping mechanism can be derived by analyzing both components (gas and dust), but such a complex analysis is out of the scope of this paper. For a more detailed modeling of the effects of dust trapping, taking into account the gas density and viscosity, we refer to \citet{sie19}.

\subsection{Origin of Ring Substructures}\label{subsec:origin}

Several physical mechanisms have been proposed to explain the presence of ring substructures in (sub)mm observations of disks. 
The most common ones can be roughly classified into three groups: planet-disk interactions \citep[e.g.][]{pap84,zhu14,bae17}, variations in the disk and/or dust properties at snowlines of volatiles \citep[e.g.][]{kre07,ros13,oku16,pin17}, and  changes in the gas dynamics/viscosity associated with the magnetic field \citep[e.g.][]{joh09,bai14,flo15,rug16}. 
Using the results of our multi-wavelength analysis, we can try to constrain the physical mechanisms responsible for the formation of the ring substructures in HD 169142. 

\subsubsection{Snowlines}

The freezeout or sublimation of gas volatiles on the surface of dust grains can significantly change the fragmentation and sticking properties of the particles \citep{gut10}. Additionally, the freezeout of volatiles near their snowlines can also result in substantial growth of the dust particles, as the amount of solid material increases \citep{ros13}. As a consequence, some studies predict the formation of annular accumulations of large particles near the snowlines \citep[e.g.,][]{pin17}. 
On the other hand, other studies have predicted that the sintering of dust grains near snowlines should increase their fragmentation rate, decreasing their size, reducing their radial migration velocity, and hence creating annular buildups of small particles that are able to reach $\tau\sim1$ \citep{oku16}. As discussed above, our results indicate that the ring substructures in HD 169142 are associated with accumulations of large dust grains, implying that, if snowlines are playing a role in HD 169142, they should be increasing the growth efficiency of dust particles.

\begin{figure}
\centering
\figurenum{8}
\includegraphics[width=0.45\textwidth]{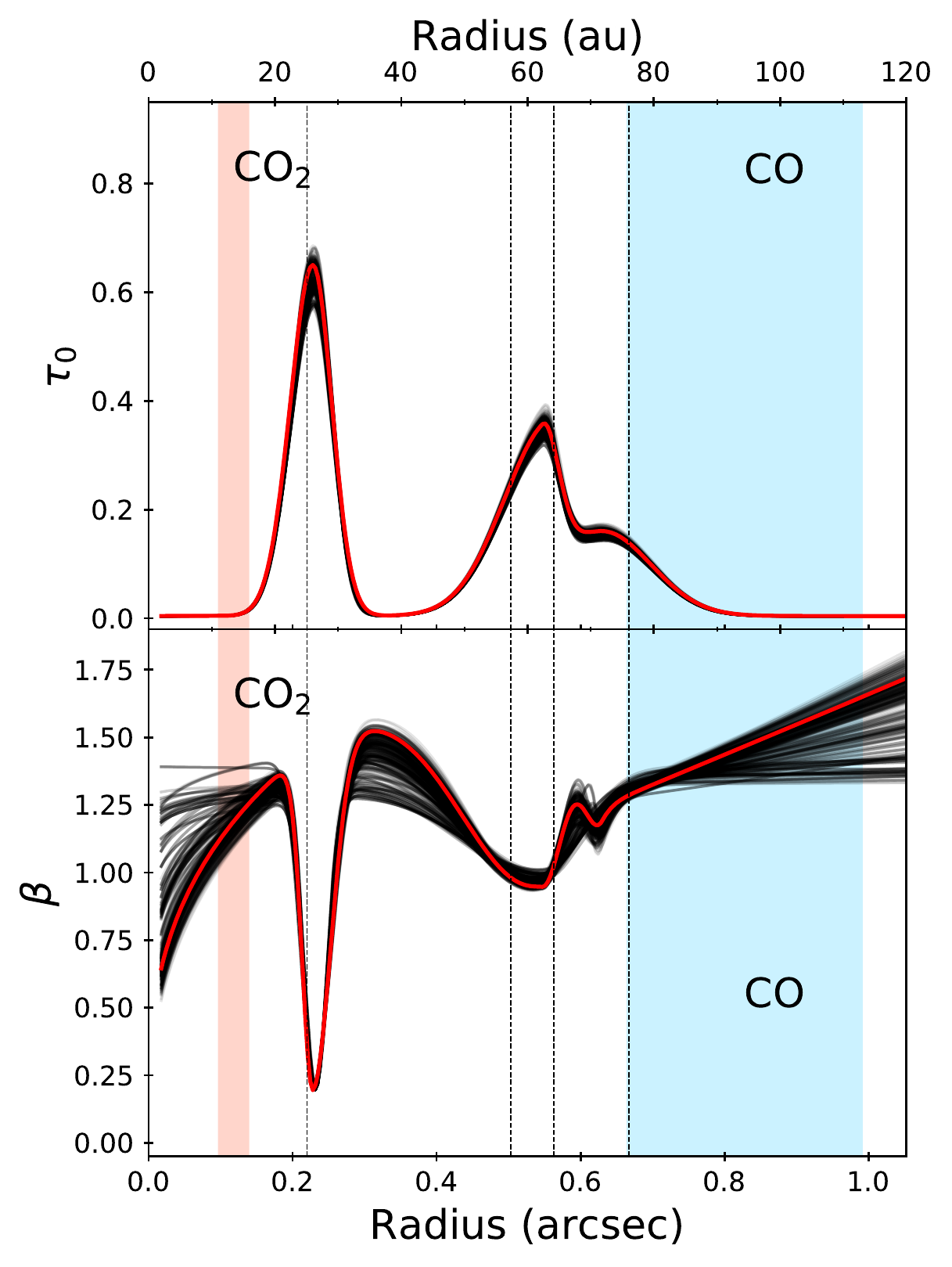}
\label{fig:snowlines}
\caption{Radial profiles of $\tau_0$ (top), and  $\beta$ (bottom) obtained from our modeling (see \S\ref{sec:modeling}), together with the range of radii where the CO and CO$_2$ snowlines are located.  The dashed vertical lines show the radius of the 4 rings reported by \citet{per19}.}
\end{figure}

We can further explore the snowline scenario by directly comparing the temperature profile obtained in our analysis to the expected position of the most important snowlines in the disk. 
We take the ranges of freezing temperature of CO$_2$, CO, N$_2$, and H$_2$O from \citet{zha15}. Their expected locations are plotted over the $\tau_0$ and $\beta$ profiles on Figure \ref{fig:snowlines}. We note that the N$_2$ snowline would fall at radii $>120$ au, whereas the H$_2$O snowline would be located between $\sim3$ and $\sim4.4$  au. Thus, these snowlines are not expected to have any effect on the observed substructures and are not plotted. 

The CO$_2$ snowline falls within the inner cavity, while the CO snowline should be located between $\sim75$ au and $\sim110$ au, beyond the B2/3 ring. The position that we predict for the CO snowline is also consistent with estimates based on the DCO$^+$ emission \citep{mac17,car18}. The snowline scenario was also tested by \citet{poh17}, who modeled VLT/SPHERE scattered light observations and estimated that the H$_2$O and CO$_2$ snowlines could be located close to B1 and B2/3. We find a slightly colder temperature profile than the one obtained by these authors, which moves the snowlines of these two volatiles to closer radii, well within the inner cavity (see Fig. \ref{fig:snowlines}).

Interestingly, given the uncertainties on the exact freezing temperatures, the CO snowline might be consistent with the position of B4, which only showed evidence of a slight increase in abundance of large dust particles. If confirmed, this could suggest that the outermost ring substructure in HD 169142 is formed through the growth of the small particles filling the outer disk as they move inward, get close to the snowline, and have their growth efficiency enhanced by the condensation of CO on their surface \citep{ros13}. As dust particles grow, they should also suffer greater drag forces that will result in a faster radial migration. However, some studies suggest that the enhanced surface density of solids at the snowlines could change the disk viscosity and trigger the formation of a gas pressure bump, which could then trap the large dust particles \citep{kre07,bit14}. 
On the other hand, another shallow ring substructure has been recently detected in near-IR polarimetric observations at $\sim90$ au and it has been suggested to be associated with the CO snowline \citep{ber18,car18}. These studies probe the small dust particles in the disk atmosphere, so it is possible that, as the disk temperature increases with height, the effects of the CO snowline are seen at a larger radii. More observations will be needed to confirm whether B4 and/or the near-IR ring at 90 au are associated with the CO snowline.

In any case, our multi-wavelength analysis indicates that at least the most prominent substructures in HD 169142 (B1 and B2/3) are not associated with snowlines. These results support recent studies of larger samples of disks, where no correlation was found between the position of the substructures and the expected position of volatile snowlines \citep{lon18,hua18b,van19}. We note that these previous studies were based on observations at a single wavelength, so they were forced to assume a certain temperature profile. By combining observations at multiple wavelengths we are able to estimate the temperature profile, obtaining a more stringent constraint on the relationship between the disk substructures and the position of the snowlines in the disk.

\subsubsection{Magnetohydrodynamic Effects}

The interaction between the magnetic field and the disk can significantly affect its gas dynamics \citep{bai17}. These magnetohydrodynamic (MHD) effects can result in the onset of radial pressure bumps that can trap large dust particles and form annular ring substructures. In general, it is difficult to distinguish between MHD effects and other mechanisms associated with pressure bumps such as planet-disk interactions \citep{rug16}, but a few key characteristics of MHD effects can be identified.

The onset of zonal flows due to the magneto rotational instability (MRI) turbulence \citep{joh09,bai14} can produce radial changes in the gas density, but with low amplitudes ($\sim10\%-20\%$) that are unlikely to induce such strong substructures as the ones in HD 169142 \citep{sim14}. On the other hand, the transition in disk ionization at the edge of the \textit{dead zone}, a region of the disk mid-plane with a lower ionization fraction, can result in a significant change in disk viscosity and, hence, in the formation of an annular pressure bump \citep{flo15,rug16}. Models usually predict this ring to form between $\sim50$ and $\sim80$ au, which could be consistent with the position of the outer rings. However, the observed triple-ring morphology \citep{per19} implies that at least two of the rings have a different origin. In particular, the narrow and compact rings B2 and B3 would be hard to reconcile with the dead-zone mechanism. More importantly, the predicted dust trap in this scenario is expected to produce azimuthal asymmetries in the form of vortices \citep{rug16}, which is inconsistent with the axisymmetry displayed by the rings in the outer disk.

Overall, MHD effects appear to be unable to explain the ring substructures in HD 169142, similarly to what has been found in other studies \citep{hua18b}. In fact, these mechanisms are predicted to be more effective in younger disks, since the magnetic field in a $\sim10$ Myr source such as HD 169142 should have been mostly removed \citep{bai17}. However, there are still several uncertainties associated with the MHD effects, such as the magnitude and evolution of the magnetic field in disks \citep[e.g.,][]{bai17}, and/or the amplitude of some of theses perturbations \citep{sim14}. 

\subsubsection{Planet-Disk Interactions}

The most commonly proposed scenario to explain the ring substructures in HD 169142 is the gravitational interaction between the disk and two or more giant planets \citep[e.g.,][]{oso14,reg14}. Based on $\sim0\rlap.''25$ resolution 1.3 mm and CO(2--1) ALMA observations, \citet{fed17} suggested the presence of a $\lesssim1$ M$_J$ planet in D1 and a $\sim1-10$ M$_J$ planet in D2, similar to what other studies found based on the gap profiles \citep{kan15,don17}. Using dust evolution models, \citet{poh17} reproduced the ring positions in their VLT/SPHERE-IRDIS and previous 1.3 mm images with a 3.5 M$_J$ planet in D1, and a second $0.7$ M$_J$ planet in D2, but these authors were  unable to fit the shallow depth of the D2 gap in scattered light. On the other hand, \citet{ber18} performed hydrodynamical simulations and successfully reproduced new SPHERE-ZIMPOL observations and ALMA archival 0.89 mm data with a 10 M$_J$ planet in D1 and two 1 M$_J$ planets in D2. One of the latter planets (at 35 au) is consistent with a blob recently revealed in SPHERE-IFS observations \citep{gra19}. These observations were not sensitive at the radial position of the second planet in D2.

All these studies were based on the assumption that the (sub-)mm rings were produced by radial dust traps on pressure bumps, but there was no robust evidence to support this. Our multi-wavelength analysis indicates that B1 and B2/3 are associated with annular accumulations of large dust particles, strongly supporting that the disk of HD 169142 harbors multiple giant planets that  are disrupting the disk.

Furthermore, \citet{per19} recently reported the triple-ring morphology of the outer disk of HD 169142 and proposed that a 10 M$_{\bigoplus}$ planet at the position of B3 could be responsible for the formation of B2, B3, and B4. This mini-Neptune would create three annular pressure bumps (one at the radius of its orbit, one at shorter radii, and one at longer radii, \citealp{don17,don18}) that would in turn trap the large dust particles of the disk forming three rings at (sub)mm wavelengths \citep{per19}. Even though we are unable to resolve the rings B2 and B3, this scenario could be consistent with our analysis, since we see evidence of dust trapping in B2/3 as well as in B4. However, according to the model by \citet{per19}, B4 should be associated with a tight and prominent dust trap, which appears to be inconsistent with our results. In fact, this dust trap also overestimates the emission of B4 in the 1.3 mm observations presented by these authors. Instead, our analysis suggests that B4 is associated with a faint accumulation of particles, that might even have an origin not related with planets. Overall, higher spatial resolution observations at 2-3 mm will be needed to accurately analyze the dust traps in the disk and discern the architecture of the planet(s) possibly responsible for this triple ring.

\section{Summary and Conclusions}

We have presented a multi-wavelength analysis of the multi-ring protoplanetary disk of HD 169142. We have reported new ALMA observations at 3 mm, which we have analyzed together with archival ALMA data at 0.89 mm and 1.3 mm. The observations at the three bands clearly resolve the characteristic double ring morphology of HD 169142, as well as some signs of the small scale substructure recently revealed in the outer disk ($r\gtrsim60$ au) by high angular resolution observations. The spectral index map between 0.89 mm and 1.3 mm shows higher values in the gaps, while lower values are found in the inner ring and the outer disk. This behavior has two possible origins that could take place at the same time: either larger particles are being accumulated in the ring substructures, and/or these substructures have optical depths close to 1.

In order to understand the origin of the changes in spectral index in the disk, we have modeled the observed visibilities using a simple axisymmetric analytical model, which yields a radial profile for the dust temperature ($T_d$), the reference optical depth at 345 GHz ($\tau_0$), and for the power of the dust opacity law ($\beta$; $\kappa_{\nu}\propto \nu ^{\beta}$). From these results we have then estimated the dust surface density and particle size distribution in the disk. The results of our analysis strongly indicate that the ring substructures in HD 169142 are the result of buildups or accumulations of large dust particles. This represents the first unambiguous evidence of the association of these ring substructures with such accumulations in HD 169142. Furthermore, we find evidence of a particularly strong and narrow buildup of large particles in the inner ring of the disk at $\sim26$ au (B1), where the conditions could be suitable enough to trigger (or have already triggered) the streaming instability. We estimate a total dust mass in the disk of $0.5^{+0.8}_{-0.3}$ M$_J$, which would represent a dust-to-gas mass ratio of $0.03^{+0.04}_{-0.02}$ if we assume the gas mass of 0.19 $M_J$ estimated by \citet{fed17}, hinting at higher ratios than the usual assumption of 0.01.

We explore different origins for the formation of the annular substructures in HD 169142. Using the results from our model we can discard dust sintering as having an important effect, since it is incompatible with the buildups of large dust particles that we find in the rings. Other mechanisms linked to the snowlines of volatiles might be associated with the outermost ring (B4), which may be located close to the CO snowline. However, other important snowlines do not appear to coincide with any substructure, pointing toward other origins associated with dust trapping at gas pressure bumps. Even though we cannot completely discard MHD effects as the origin of the dust traps in the outer disk, the age of HD 169142, as well as the narrow width and high contrast of the rings, make this scenario unlikely. Overall, our results strongly support the planet origin scenario, in agreement with other recent studies on HD 169142.

Multi-wavelength studies represent the most powerful tool to analyze the size distribution of dust particles in the disk and its association with disk substructures. Extending this study to a larger sample of objects will allow us to confirm whether disk substructures can trap large dust particles, and provide a suitable environment for planetesimal formation.

\acknowledgments

E.M., and C.C.E. acknowledge support from the National Science Foundation under CAREER Grant Number AST-1455042 and the Sloan Foundation. 
M.O., G.A., J.M.T., and J.F.G. acknowledge financial support from the State Agency for Research of the Spanish MCIU through the AYA2017-84390-C2-1-R grant (co-funded by FEDER). M.O., G.A., and J.F.G. acknowledge support from the ``Center of Excellence Severo Ochoa'' award for the Instituto de Astrof\'{\i}sica de Andaluc\'{\i}a (SEV-2017-0709).
M.F. and G.H.M.B. acknowledge support from the European Research Council (ERC) under the European Union's Horizon 2020 research and innovation programme (gran agreement n$^{\circ}$ 757957).

This paper makes use of the following ALMA data: ADS/JAO.ALMA\#2012.1.00799.S, JAO.ALMA\#2015.1.00490.S, JAO.ALMA\#2015.1.01301.S, and JAO.ALMA\#2016.1.01158.S. ALMA is a partnership of ESO (representing its member states), NSF (USA) and NINS (Japan), together with NRC (Canada), MOST and ASIAA (Taiwan), and KASI (Republic of Korea), in cooperation with the Republic of Chile. The Joint ALMA Observatory is operated by ESO, AUI/NRAO and NAOJ.
This work has made use of data from the European Space Agency (ESA) mission {\it Gaia} (\url{https://www.cosmos.esa.int/gaia}), processed by the {\it Gaia} Data Processing and Analysis Consortium (DPAC, \url{https://www.cosmos.esa.int/web/gaia/dpac/consortium}). Funding for the DPAC has been provided by national institutions, in particular the institutions participating in the {\it Gaia} Multilateral Agreement. 



\facilities{VLA, ALMA}

\software{\texttt{CASA} (v5.3.0; \citep{mcm07}), \texttt{EMCEE} \citep{for13}, \texttt{Numpy} \citep{van11}, \texttt{Matplotlib} \citep{hun07}, \texttt{Astropy} \citep{ast18}}

\clearpage


\listofchanges

\end{document}